\documentclass[12pt]{article}
\usepackage[T1]{fontenc}
\usepackage[utf8]{inputenc}
\usepackage{lmodern}
\usepackage{amsmath}
\usepackage{amssymb}
\usepackage{graphicx}
\title{Induced rotation from de-Sitter--G\"odel--de-Sitter phase transition}
\author{Sh. Khodabakhshi$^{1}$, A. Shojai$^{1,2}$ \\
$^1$Department of Physics, University of Tehran,\\
$^2$Foundations of Physics Group, School of Physics,\\ Institute for Research in Fundamental Sciences (IPM).}
\date{}
\begin{document}
\maketitle
\begin{abstract}
Rotation of the cosmic objects is a universal phenomenon and it's origin is still an open question. Here a model for the origin of rotation is presented. After investigation of the phase transition of a scalar field in de-Sitter and G\"odel backgrounds, the motion of a test particle during the phase transitions is studied. Then using computer simulation for a congruence of particles, we have shown that although the local induced rotation is non--zero, the global rotation is below the observational limit.
\end{abstract}
\section{Introduction}
All cosmic objects from stars to the clusters of galaxies are rotating. Rotation should be an essential property of any theory for the origin of galaxies. Rotation prevents  celestial bodies from collapsing. If there were no rotation, intuitively it would seem like everything would start falling directly toward the nearest largest mass. As stated by Whittaker in his lecture \cite{1} : “Rotation is a universal phenomenon; the earth and all the other members of the solar system rotate on their axes, the satellites revolve round the planets, the planets revolve round the Sun, and the Sun himself is a member of the galaxy or Milky Way system which revolves in a very remarkable way. How did all these rotary motions come into being? What secures their permanence or brings about their modifications? And what part do they play in the system of the world?”. 

Here particular attention is paid to Whittaker's question. Although many attempts were made to explain how the rotation got started, there is no unanimity on this topic.       Weizs\"acker \cite{4}, Ozernoy and Chernin \cite{5,6}, developed the scenario of primordial turbulences. Now it is only of historical value because if the angular momentum of galaxies had originated by turbulence, their spins should be perpendicular to the main proto-structure plane \cite{7} which is not observed. Also if the rotation of spiral galaxies were remained from the primordial turbulence, it would be dissipated by shocks at the epoch of recombinations \cite{8}. 

Further possible process of explaining the origin of rotation is vector perturbation of the metric. One possible source for such perturbations is the cosmic strings but no convincing experimental verification of cosmic strings has been confirmed up to now \cite{2,3}. One may also follow the idea of Hoyle and say that the origin of rotation would have come from angular momenta of its original neighbors. However, it is difficult to prove it in a  theoretical treatment \cite{13}.

Of particular interest is the idea of rotation of the  whole universe. This idea was first expressed by G. Gamow \cite{9}. Soon after Gamov's letter, G\"odel derived such a solution for Einstein's field equations on his seventieth birthday. (for a brief history see \cite{10}). If we say that the universe is rotating as a whole we would have to define a rotation axis. A rotation axis would cause some problems, like violating a homogeneous and isotropic universe as proposed by the standard model of cosmology and also is in contrast with the  cosmological microwave background radiation (CMBR). G\"odel also admitted that his universe can not serve as a model of the universe we live in because it does not contain expansion. 

Although the attempts mentioned above are valued, Whittaker's question is still unanswered. It is necessary to achieve a mechanism based on standard theories of physics i.e. quantum field theory and the standard model of cosmology for the origin of rotation which is consistent with the observational data.
A possible scenario is that the universe make a phase transition from the standard cosmological solution to a rotating model and then to the standard cosmological solution. This phase transition can be caused by a scalar field.

 The overview \cite{12} consists of the list of references on cosmological models with rotation but perhaps the G\"odel space-time is the best known of them  \cite{11}. It was suggested in \cite{14} that a phase transition between the Friedmann open metric and G\"odel  can lead to a new cosmological scenario for the evolution of the universe, in which the present universe is originated from primordial G\"odel universe. The cosmological phase transition in G\"odel space--time using the zeta function method has been studied in \cite{15} and also the effective action of a scalar field in G\"odel space--time was obtained in \cite{16}. 

In this paper we adopt a scenario in which the universe experiences a phase transition from de-Sitter  to G\"odel space--time and then from G\"odel to de-Sitter space--time. We show that such a double transition would induce rotation into the trajectory of initially non--rotating test particles. Since in this scenario the universe lives in G\"odel phase for a small time, it has the benefit that the rotation is introduced without destroying other aspects of the cosmological model. Although for simplicity, we studied here the phase transition from static de-Sitter to G\"odel and then back to de-Sitter space--time, the scenario is applicable to more realistic cases.

This paper is organized as follows: in the next section, we derive the effective potential of a scalar field in de-Sitter and G\"odel space--times in a suitable form for our scenario. Then in section three, the geodesics of a test particle is investigated during the phase transition and the induced rotation is obtained. Finally, using computer simulation, the local and global induced rotation for a congruence of particles is studied.

\section{de-Sitter--G\"odel--de-Sitter phase transition}
Let's start with a scalar field living on a curved background. Comparing covariant expression for the components of the energy--momentum tensor of a perfect fluid
and the energy--momentum tensor of a scalar field
makes it clear that a scalar field acts like a perfect fluid with an energy density and pressure given by
\begin{equation}
p_{\phi}=\frac{1}{2}\dot{\phi}^{2}-V(\phi)-\frac{1}{2}\left|\vec{\nabla}\phi\right|^{2}
\end{equation} 
\begin{equation}
\rho_{\phi }=\frac{1}{2}\dot{\phi}^{2}+V(\phi)+\frac{1}{2}\left|\vec{\nabla}\phi\right|^{2}
\end{equation} 
In the minimum of potential, $ \dot{\phi} $ is negligible and $ \vec{\nabla}\phi=0 $ provided that the field $ \phi  $ is spatially constant. These assumptions yield an equation of state of $ p_{\phi}=-\rho _{\phi} $, thus $ V(\phi) $ acts as an effective cosmological constant. To consider the quantum effects we have to employ the  method of the effective potential in curved space--times \cite{17,18}. Given the action of a scalar field on a curved background
\begin{equation}
S[\phi,g_{\mu \nu }]=\int d^{4}x\sqrt{-g}\left(\frac{1}{2}\partial_{\mu}\phi\partial^{\mu}\phi-V(\phi)\right)
\end{equation}
the expression of one-loop effective potential is
\begin{equation}
V^{(1)}_{\textrm{eff}}=-\frac{\Gamma^{(1)}}{\cal{V}}
\end{equation}
where $ \Gamma^{(1)}=-\frac{i}{2}\ln(\mu ^{-2}\det G) $ is the  one-loop effective action, $G=\square +V''(\phi)$, and $\cal{V}$ is the spatial volume. The arbitrary length parameter $ \mu  $ is introduced for dimensional considerations and $ \square $ is the Laplace operator defined with the background metric.

Since the universe is in thermal equilibrium, computations should be performed in the framework of finite temperature quantum field theory hence we use the Euclidean effective potential. It is important to notice that the wick rotation in curved space--time is a bit problematic. We can only adopt wick rotation for stationary space--times. Also, for rotating metrics, more parameters than the time coordinate should be analytically continued. 

Here we are looking for a situation in which the space--time metric changes from de-Sitter to G\"odel and vice versa as a result of thermal phase transition of the scalar field. 

\begin{figure}
  \begin{center}
    \includegraphics[height=15cm]{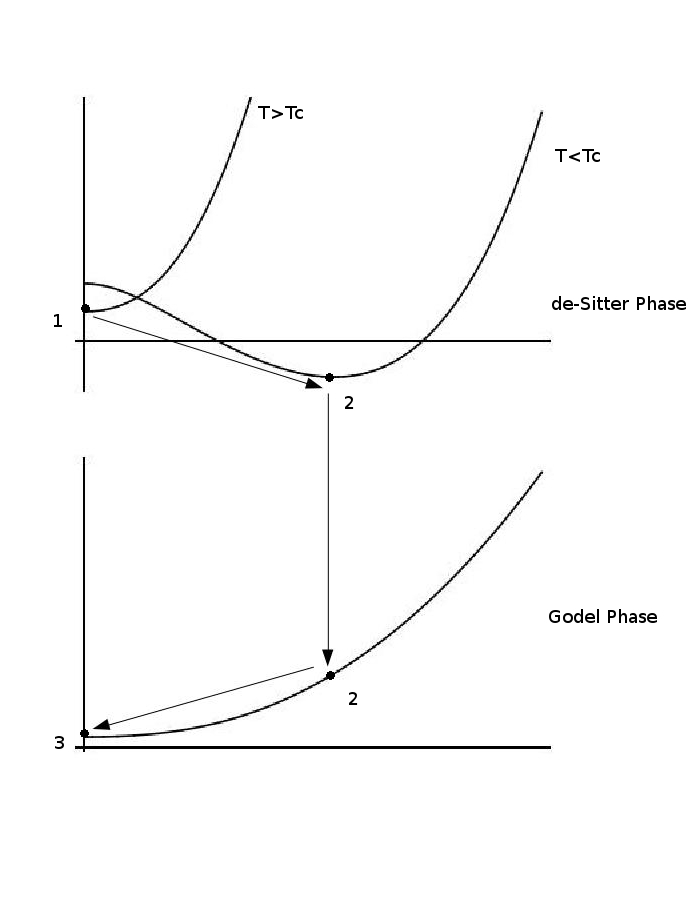}
    \caption{A typical scenario of the universe going through de-Sitter, G\"odel, and de-Sitter phases.}
       \label{fig: 1}
  \end{center}
\end{figure}

In order to make the mentioned scenario working, we have to have three phases. First, $V_{\textrm{eff}}$ acts like a positive cosmological constant which is much larger than the dust density ($V_{\textrm{eff}}\gg \rho_{\textrm{dust}}$) leading to de-Sitter space--time. Second, $V_{\textrm{eff}}$ plays the role of a negative cosmological constant and equal to $-\rho_{\textrm{dust}}/2$ to have G\"odel space--time. Finally the scalar field should roll such that the universe arrives at de-Sitter space--time again.
A typical scenario of this process is shown in figure (\ref{fig: 1}). 

To continue, let's first briefly investigate the finite temperature one--loop potential of a scalar field in de-Sitter and G\"odel backgrounds.  
\subsection{de-Sitter--G\"odel phase transition}
Although de-Sitter solution of the Einstein's equations is a non-stationary space--time, we restrict ourselves to the static chart to have an stationary metric in which the wick rotation can be applied. The  Euclidean line element of de-Sitter space--time in the static coordinate patch is:
\begin{equation}
 ds^{2}=\cos^{2}\chi d\tau^{2}+a^{2}(d\chi ^{2}+\sin^{2}\chi d\theta ^{2}+\sin^{2}\chi\sin^{2}\theta d\xi^{2})
 \label{ds}
 \end{equation} 
where $ -\infty<t=-i\tau<+\infty $, $ -\pi<\chi<\pi $, $ 0<\theta,\xi<\pi $, $ a^2=\frac{2}{\Lambda}$ is the spatial scale, and the periodic parameter $ \tau $ ranges from $ 0 $ to $ \beta $\cite{19}.
Using the zeta-function regularization method, one can derive the effective potential as
\begin{equation}
 V_{\textrm{eff}}(\phi ,\beta )=V(\phi)-\frac{1}{2\beta \cal V}\left [\zeta'(0,\beta)+\log(\mu^{2}a^{2})\zeta(0,\beta)+\log(V''(\phi)\mu^{-2})\right ]
 \end{equation} 
 To be specific, we assume that the classical potential is:
 \begin{equation}
 V=\frac{1}{2}\sigma^2\phi^2+\frac{1}{24}\lambda\phi^4
 \end{equation}
Following the calculations of Appendix A, the renormalized effective potential can be written in terms of dimension-less quantities as:
\[ \frac{\lambda}{2\sigma^4}V_{\textrm{eff}}=\frac{\lambda}{2\sigma^4}V_0+
\]
\[
\frac{1}{2}x^2+\frac{1}{12}x^4-\gamma T\Bigg\{ 0.62\Delta-0.22\Delta^2+0.075 T+\frac{T^3}{120}+\frac{1}{T}\Big (-0.07+2.7\Delta^2-0.15\Delta\Big)
\]
\begin{equation}
+\Big(0.22-\frac{2.7}{T}\Big)\Big(\frac{51-60T^2-8T^4}{240}-\frac{2T^2-3}{2}\Delta+\Delta^2\Big)
+\log[\alpha(1+x^2)]\Bigg\}
\label{vd}
\end{equation}
where $\alpha=a^2\sigma^2$, $\gamma=\frac{3\lambda}{32\pi^2\alpha^2}$, $x=\sqrt{\frac{\lambda}{2}}\frac{\phi}{\sigma}$, $\Delta=\frac{9}{4}-\alpha(1+x^2)$, and $T=\frac{\beta _{H}}{\beta}=\frac{2\pi a}{\beta}$.

\begin{figure}
  \begin{center}
    \includegraphics[height=15cm]{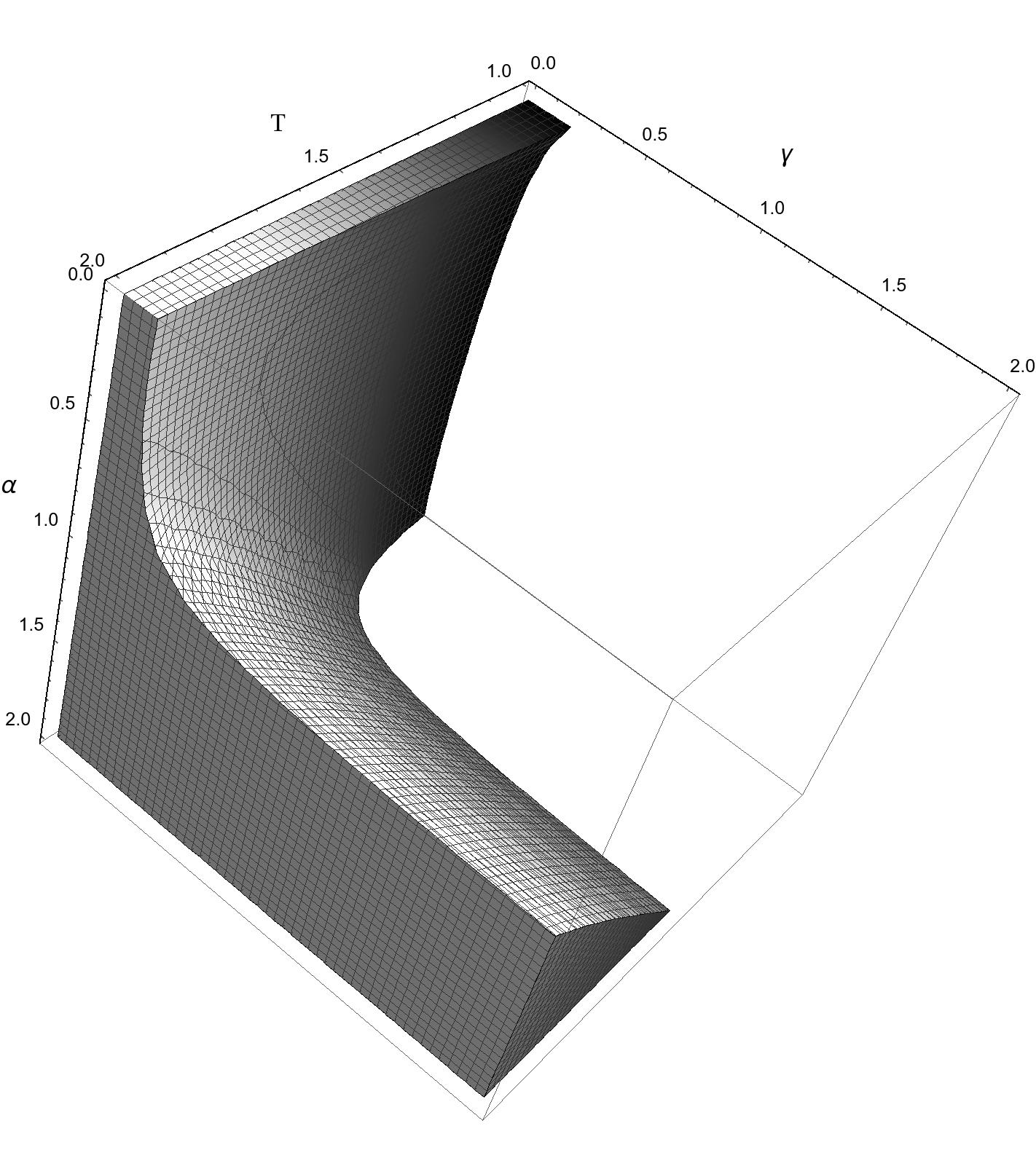}
    \caption{Regions that the potential has a Mexican hat shape.}
       \label{fig: 2}
  \end{center}
\end{figure}

This potential has extrema at: 
\begin{equation}
x=0
\end{equation}
and at the roots of the equation:
\[
1+\frac{1}{3}x^3-\gamma T\Bigg\{ -1.24\alpha +0.44\Delta-\frac{10.8}{T}\Delta+\frac{0.3\alpha}{T}+
\]
\begin{equation}
\alpha\left (0.22-\frac{2.7}{T}\right)(2T^2-3-4\Delta)+\frac{2}{1+x^2}\Bigg\}=0
\label{vd1}
\end{equation}
Note that if the above equation has no real root, the potential is a harmonic--type potential and otherwise it is in Mexican hat shape. The regions of parameters $\alpha$, $\gamma$, and the temperature $T$ that allows a Mexican hat shape is plotted in figure (\ref{fig: 2}). 

\begin{figure}
  \begin{center}
    \includegraphics[height=7cm]{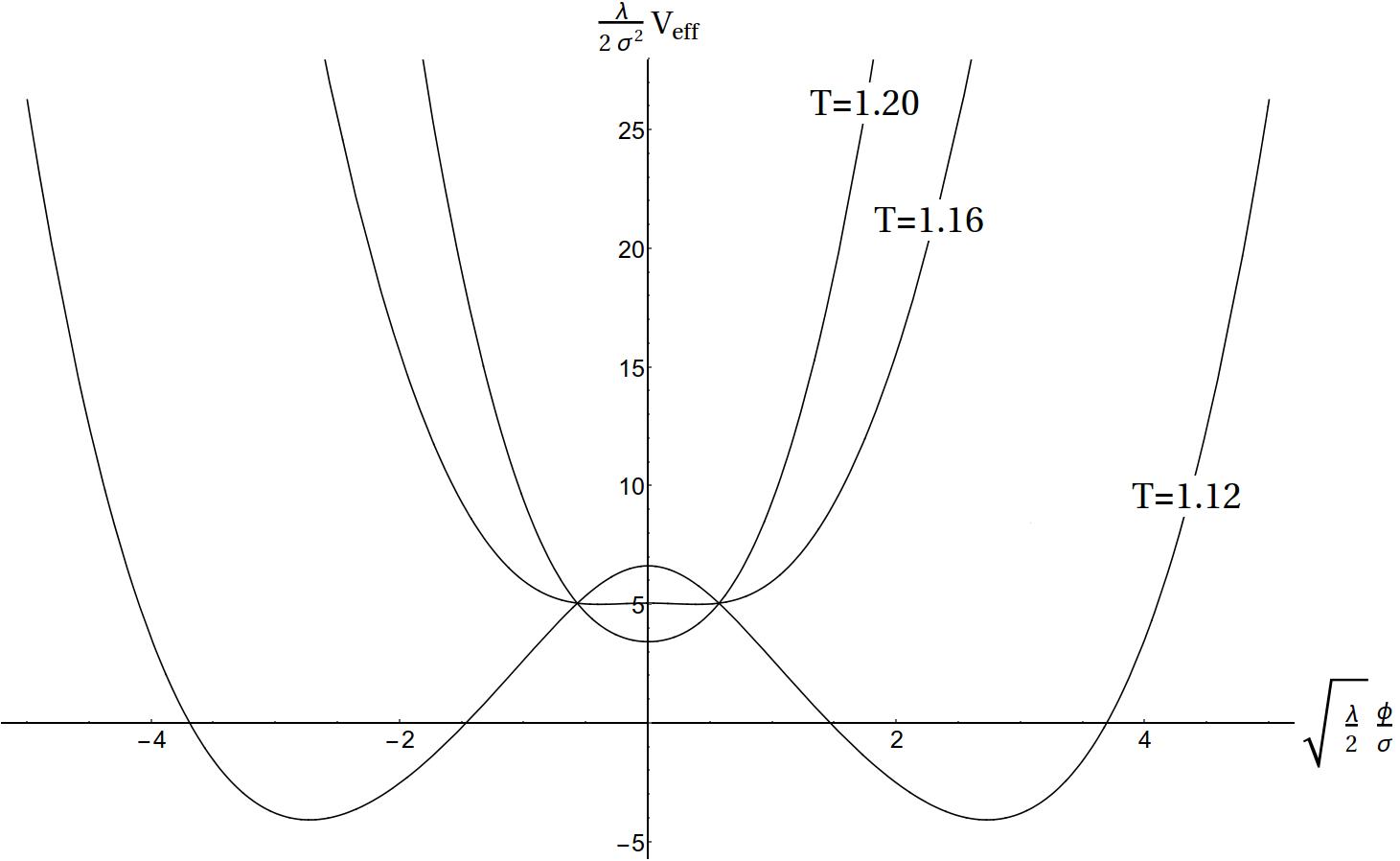}
    \caption{Typical shape of the one--loop potential for different values of temperature, $\alpha=2$ and $\gamma=10$.}
       \label{fig: 3}
  \end{center}
\end{figure}

It is clear that during the cooling of the universe, there is a critical temperature that the scalar field experiences a  phase transition as desired. This is plotted in figure (\ref{fig: 3}), for some specific values of the parameters.
At the temperature that the minimum of the Mexican hat has depth equal to $-\Lambda=\rho_{\textrm{dust}}/2$, the universe experiences G\"odel geometry and thus we have to evaluate the one--loop potential in G\"odel's background, as we do in the next subsection.
\subsection{G\"odel--de-Sitter phase transition}
G\"odel space-time is an exact solution of Einstein's equations with a negative cosmological constant. The source for G\"odel geometry is a perfect fluid with a constant density $ \rho $ and no pressure $ p=0 $ . In the Cartesian coordinates the G\"odel line element is given by
\begin{equation}
ds^{2}=(dt+e^{\sqrt{2}\Omega x/2}dy)^{2}-dx^{2}-\frac{1}{2}e^{\sqrt{2}\Omega x}dy^{2}-dz^{2}
\end{equation}
where the constant $\Omega$ determines the angular momentum four-vector ($\Omega ^{\beta}=(0,0,0,\Omega)$) of the intrinsic rotation of the whole manifold.

Following calculations of \cite{15}, the effective potential in G\"odel background for a massive scalar field is given by
\[
V_{\textrm{eff}}=\bar V_0+ \frac{\sigma^4}{16\pi ^{3/2}}\left(\frac{\Omega^2}{8\sigma^2}+\frac{M^2}{\sigma^2}\right)^{3/2}\times
\]
\begin{equation}
\left(\frac{2\sqrt{\pi}}{3}-\sum_{n,\ell }^{'}\left [2Z^{\nu-3/2}_{1}K_{3/2-\nu}(2Z_1)-Z^{\nu-3/2}_{2}K_{3/2-\nu}(2Z_2)\right]\right)
\end{equation}
where $ M^2=\sigma^2+\frac{1}{2}\lambda\phi^2$, and $\bar V_0$ is a constant. Prime on the summation denotes that the term $ n=\ell=0 $ is omitted. $ K $ is the modified Bessel function and $ Z $ is the Epstein zeta function 
\begin{equation}
Z_{1}=\pi\sqrt{\left(\frac{\Omega^2}{8}+M^{2}\right)\left(\frac{\ell^{2}}{4\pi^{2}\bar T^{2}\sigma^2}+\frac{8n^{2}}{\Omega^{2}}\right)}
\end{equation}
\begin{equation}
Z_{2}=\pi\sqrt{\left(\frac{\Omega^2}{8}+M^{2}\right)\left(\frac{\ell^{2}}{4\pi^{2}\bar T^{2}\sigma^2}+\frac{2n^{2}}{\Omega^{2}}\right)}
\end{equation}
in which $\bar T^{-1}=\beta\sigma$. In the limit $\bar\alpha=8\sigma^2/\Omega^2\gg 1 $ and $ \bar T\gg 1$, the renormalized effective potential as a function of dimension-less quantities is given by
\begin{equation}
\frac{\lambda}{2\sigma^4}V_{\textrm{eff}}=\frac{\lambda}{2\sigma^4}\bar V_0-\bar\gamma \bar T \left(1+\bar\alpha(1+x^2)\right)^{1/4}\exp\left [-\pi\sqrt{1+\bar\alpha(1+x^2)}\right ]
\label{vg}
\end{equation}
where $\bar\gamma=\frac{\lambda\Omega^2}{16\pi^2\sqrt{2}\sigma^2}$.

\begin{figure}
  \begin{center}
    \includegraphics[height=7cm]{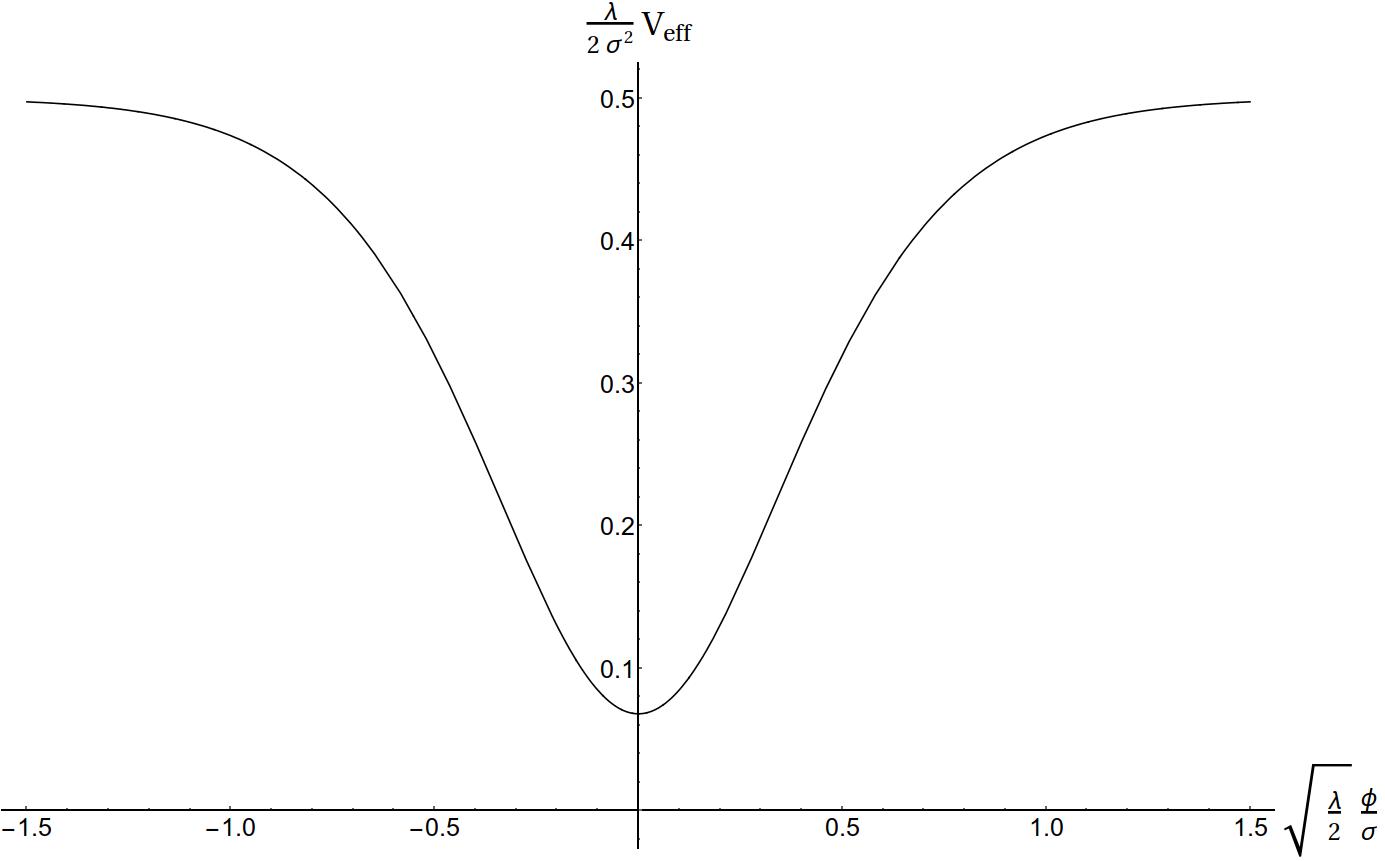}
    \caption{One--loop efective potential in G\"odel's background with $\bar\alpha=3$, $\bar\gamma\bar T=10$, and $\frac{\lambda V_0}{2\sigma^2}=0.5$.}
       \label{fig: 4}
  \end{center}
\end{figure}

The general shape of the on--loop effective potential in these limits is shown in figure (\ref{fig: 4}).

Now we have enough material to bring our desired scenario into act. Starting from a de-Sitter background (with $V_0\sim\Lambda$), after enough cooling of the universe via expansion, the universe reaches the critical temperature $\beta_c$. This temperature is where $\left. \frac{d^2V_{\textrm{eff}}}{d\phi^2}\right|_{\phi=0}=0$. Assuming $\lambda$ is small, from the equation (\ref{vd}), one can see that the approximate relation for the critical temperature is:
\begin{equation}
\frac{2\pi a}{\beta_c}\simeq 4.1+(-196.5+212.8\alpha)\gamma^{-1/3}+\cdots
\end{equation}
Below this temperature, the scalar field do a phase transition to the Mexican hat potential (see figures (\ref{fig: 1}) and (\ref{fig: 3})). 
The phase transition can be studied using the mean field approximation. It is a second order phase transition with the free energy given by 
\begin{equation}
e^{-\beta F}=\int d\phi\ e^{-\beta\int d^3x\sqrt{-g} V_{\textrm{eff}}}
\end{equation}
 and the order parameter $\langle\phi\rangle$. From equation (\ref{vd}) we see that:
\begin{equation}
{\cal F}=\int d^3x\sqrt{-g} V_{\textrm{eff}}={\cal F}_0+A(T)\phi^2+\frac{\lambda{\cal V}}{24}\phi^4+\cdots
\end{equation}
where ${\cal F}_0$ is a constant and $A(T)$ has a Taylor expansion $A(T)=\sum a_n \left(\frac{T-T_c}{T_c}\right)^n$. The critical exponents can be simply obtained as:
\begin{equation}
\langle\phi\rangle\sim\left(\frac{T-T_c}{T_c}\right)^{1/2}
\end{equation}
\begin{equation}
\langle\phi^2\rangle-\langle\phi\rangle^2\sim\left(\frac{T-T_c}{T_c}\right)^{-1}
\end{equation}
and the critical exponent of the specific heat is zero.

The mean field can be seen from equation (\ref{vd1}) that is approximately $\phi\sim\sigma/\sqrt{\lambda}$. In order to have a G\"odel space--time, the depth of potential should be half the density of the matter. Using equation (\ref{vd}), we have $\rho_{\textrm{dust}}\sim\sigma^2/\lambda$. 
Since the space--time metric changes from de-Sitter to G\"odel, the scalar field potential changes shape to figure (\ref{fig: 4}).

In the G\"odel phase, the scalar field rolls down the potential till $\phi=0$. In order to not introduce any changes in the cosmological parameters, we set $\bar V_0\simeq \Lambda$. Using equation (\ref{vg}), we see that the universe again goes to a de-Sitter phase, after elapsing the time given by:
\begin{equation}
\tilde t\simeq\frac{\phi}{\dot\phi}\simeq\frac{\sigma}{\sqrt{\lambda\Lambda}}
\end{equation}

Writing anything in terms of the more physical parameters $\Lambda$, $\rho_{\textrm{dust}}$, and $\beta$, and recovering $c$, $\hbar$ and $G$ we have:
\begin{equation}
\left \{
\begin{matrix}
V_0=\Lambda&,&\\
\bar V_0=\Lambda&,&\\
\lambda=\frac{\Lambda^{7/8}}{(\hbar\beta c)^{3/4}}\left ( \frac{G\rho_{\textrm{dust}}}{c^2}\right )^{-5/4}&,&\\
\sigma^8=\frac{\Lambda^{7/2}}{(\hbar\beta c)^{3}}\left ( \frac{G\rho_{\textrm{dust}}}{c^2}\right )^{-1}&,&\\
\tilde t=\sqrt{\frac{\hbar G^2\rho_{\textrm{dust}}}{\Lambda c^7}}&.&\\
\end{matrix}
\right .
\label{roll}
\end{equation}
It is important to note that since we assumed that the dust density is very smaller than the cosmological constant, the time that universe elapses in G\"odel phase is much smaller than the time scale of de-Sitter space--time.

At this end, some notes on the horizon temperature is useful. The horizon temperature of de-Sitter space is defined as $\beta_H=2\pi a=2\sqrt 2\pi/\sqrt\Lambda$. In our simple scenario, we assumed that the space--time after leaving the G\"odel phase, enters a de-Sitter phase with the same cosmological constant as the first de-Sitter space, and therefore the horizon temperature does not changes during the whole process. This is not an essential requirement for this model. One can assume that the cosmological constant in the first and final phases are different. This makes no essential changes in the results and thus we assumed here that no change in the horizon temperature is present. Another thing that is important to note is that the critical temperature and the horizon temperature are not equal. For example, for the parameters given in figure (\ref{fig: 3}), the shape change is at $T\sim 1.16$, or $\beta_H=1.16\beta$. Therefore at the critical temperature, the horizon and the scalar field are not at thermal equilibrium. But as it is clear from  (\ref{fig: 3}), transition to a true vacuum of the Mexican hat potential, brings these more closer to equilibrium.   

\section{Motion of test particles during the phase transition}
Now we have a model in which the background goes through de-Sitter to G\"odel and then to de-Sitter space--times. this is a result of the phase transition and slow rolling of a scalar field. We are now ready to investigate the motion of test particles during these phase transitions. It is quite reasonable to expect that a particle with a non-rotating trajectory entering the first de-Sitter phase, exits to the final de-Sitter phase while it is rotating. This leads to an induced rotation which could be considered as the reason of universal rotation.

In order to see this, in what follows, we start from a particle moving radially  and match its trajectory to the G\"odel's and then to de-Sitter's  geodesics.
Consider a massive particle which initiates at the point $ (t_{1},r_{1},\frac{\pi}{2},0) $ in the first de-Sitter phase (point number 1 in figure (\ref{fig: 1})). The de-Sitter static patch line element in the coordinates $ (t,r,\theta ,\phi  ) $ is given by:
\begin{equation}
ds^{2}=\left( 1-\frac{r^{2}}{a^{2}}\right )dt^{2}-\left(1-\frac{r^{2}}{a^{2}}\right)^{-1}dr^{2}-r^{2}\left(d\theta ^{2}+\sin^{2}\theta d\phi ^{2}\right )
\end{equation}
The four--velocity of the particle at point number 1, is given by:
\begin{equation}
U^{\mu(D_{1})}=\left (U^0,U^1,0,0\right )
\end{equation}
where $D_{1}$ superscript means that the particle is in the first de-Sitter phase.
Now the scalar field suddenly makes a phase transition to the point number 2 of figure (\ref{fig: 1}), and thus the space--time metric becomes G\"odel metric. In order to find the initial condition of the particle, we have to match these phases. We do this by writing the above initial condition in the local inertial frame and then converting to the G\"odel's global coordinate system. To do so we need the tetrads of both space--times. 

de-Sitter's tetrads are given by:
\begin{equation}
e^{a}_{\mu}=
            \begin{pmatrix}
              \left (1-\frac{r^2}{a^2}\right)^{\frac{1}{2}}& 0 & 0 & 0\\
              0 & \left(1-\frac{r^2}{a^2}\right)^{-\frac{1}{2}} & 0 & 0\\
              0 & 0 & r & 0\\
              0 & 0 & 0 & r\sin \theta
            \end{pmatrix}
\end{equation}
and
 \begin{equation}
e^{\mu}_{a}=
            \begin{pmatrix}
              \left(1-\frac{r^2}{a^2}\right)^{-\frac{1}{2}} & 0 & 0 & 0\\
              0 & \left(1-\frac{r^2}{a^2}\right)^{\frac{1}{2}} & 0 & 0\\
              0 & 0 & r^{-1} & 0\\
              0 & 0 & 0 & (r\sin \theta)^{-1}
            \end{pmatrix}
\end{equation}
 
For G\"odel's space--time, it is better to write the  line element in cylindrical coordinates $(t,r,\phi,z)$:
\begin{equation}
ds^2=\frac{2}{\Omega ^{2}}\left[dt^2-dr^2-dz^2+(\sinh^4r-\sinh^2r)d\phi^2+2\sqrt{2}\sinh^2r\ d\phi dt\right]
\end{equation} 
where $2\Omega^2=\rho_{\textrm{dust}}$. The following change of coordinates makes it more suitable for our purpose:
\begin{equation}
\frac{\sqrt 2}{\Omega}t=t',\ \ \ \ \ \ \ \frac{\sqrt 2}{\Omega}r=r',\ \ \ \ \ \ \ \frac{\sqrt 2}{\Omega}\phi=\phi',\ \ \ \ \ \ \ \frac{\sqrt 2}{\Omega}z=z'
\end{equation}
The metric then would be:
\begin{equation}
ds^2=dt'^2-dr'^2-dz'^2+(\sinh^4\xi-\sinh^2\xi)d\phi^2+2\sqrt{2}\sinh^2\xi\ d\phi' dt'
\end{equation}
where $\xi=\Omega r'/\sqrt{2}$. 

The tetrads of G\"odel's space--time in this coordinate system are then computed easily as:
\begin{equation}
e^{a}_{\mu}=
            \begin{pmatrix}
              1 & 0 & \sqrt{2}\sinh ^{2}\xi & 0\\
              0 & 1 & 0 & 0\\
              0 & 0 & \sinh\xi\cosh\xi & 0\\
              0 & 0 & 0 & 1
            \end{pmatrix}
\end{equation}
and
\begin{equation}
e^{\mu}_{a}=
            \begin{pmatrix}
              1 & 0 & -\sqrt{2}\tanh \xi & 0\\
              0 & 1 & 0 & 0\\
              0 & 0 & \left(\sinh\xi\cosh\xi\right)^{-1} & 0\\
              0 & 0 & 0 & 1
            \end{pmatrix}
\end{equation}

Now the velocity in local inertial frame of de-Sitter space is:
\begin{equation}
U^a=e^{a(D_{1})}_{\mu }U^{\mu (D_{1})}=\left( U^0\sqrt{1-r_1^2/a^2},U^1/\sqrt{1-r_1^2/a^2},0,0\right )\equiv\left( \gamma, \gamma v,0,0\right)
\end{equation}
where $v$ is the local radial velocity, and $\gamma=(1-v^2)^{-1/2}$. Therefore
 the initial velocity of the particle in the G\"odel space-time is:   
\begin{equation}
U^{\nu (G)i}=e^{\nu(G)}_{a}U^a=\left ( \gamma,\gamma v,0,0\right)
\end{equation}
During the slow roll of the scalar field from point number 2 to point number 3 of figure (\ref{fig: 1}), the particle evolves according to the geodesics equation from $ t'_{2} $ to $t'_{3}=t'_{2}+ \tilde{t} $, where $ \tilde{t} $ is given by equation (\ref{roll}). The trajectory can be derived from the geodesics equations of the G\"odel space--time:
\begin{equation}
\ddot{r'}+(2\cosh \xi\sinh^3\xi-\sinh \xi\cosh \xi)\dot{\phi' }^2+2\sqrt{2}(\sinh \xi\cosh \xi)\dot{\phi' }\dot{t'}=0
\label{eq1}
\end{equation}
\begin{equation}
\dot{t'}+\sqrt{2}(\sinh^2\xi)\dot{\phi' }=A
\label{eq2}
\end{equation}
\begin{equation}
\dot{\phi'}(\sinh^4\xi-\sinh^2\xi)+\sqrt{2}(\sinh^2\xi)\dot{t'}=B
\label{eq3}
\end{equation}
where $ A $ and $ B $ are constants. Furthermore for massive particles, we have:
\begin{equation}
\dot{t'}^2-\dot{r'}^2+(\sinh^4\xi-\sinh^2\xi)\dot{\phi' }^2+2\sqrt{2}(\sinh^2\xi)\dot{\phi' }\dot{t'}=1
\label{eq1a}
\end{equation}

Applying the  initial condition at the point $ r'_{2}$, the constants A and B are simply obtained in terms of $\xi_2=\Omega r'_{2}/\sqrt{2}$:
\begin{equation}
A=\gamma,\ \ \ \ \ \ \ 
B=\sqrt{2}\gamma \sinh^{2}\xi_2
\end{equation}

Finally from the relations (\ref{eq2}), (\ref{eq3}), and (\ref{eq1a}), one can obtain  $ \dot{t'} $, $ \dot{\phi'}$, and $\dot r'$ as:
\begin{equation}
\dot{t'}=\gamma  \left ( 1-2\frac{\sinh^2\xi-\sinh^2\xi_2}{\cosh^2\xi}\right )
\label{eq4}
\end{equation} 
\begin{equation}
\dot{\phi'}=\sqrt{2}\gamma \frac{\sinh^2\xi-\sinh^2\xi_2}{\sinh^2\xi\cosh^2\xi}
\label{eq5}
\end{equation}
\begin{equation}
\dot{r'}=\frac{\sqrt{v^2-2\frac{(\sinh^2\xi-\sinh^2\xi_2)^2}{\sinh^2\xi\cosh^2\xi}}}{1-2\frac{\sinh^2\xi-\sinh^2\xi_2}{\cosh^2\xi}}
\label{eqr5}
\end{equation}

During the field roll from point number 2 to the point number 3 of figure (\ref{fig: 1})  the velocity would be
\begin{equation}
U^{\mu(G)f}=(\dot{t}'_{3},\dot{r}'_{3},\dot{\phi}'_{3},0)
\end{equation}
In principle $ \dot{t}'_{3},\dot{\phi}'_{3} ,\dot{r}'_{3}$ can be obtained from equations (\ref{eq4}),(\ref{eq5}) and (\ref{eqr5}), at least numerically. But since the time $\tilde t$ is small compared to the time scale of G\"odel metric (note that $\Omega\tilde t=G\rho_{\textrm{dust}}/\Lambda c^2 \ll 1$), we can expand the relations around $\xi_2$ to obtain:
\begin{equation}
\frac{d\xi}{dt'}\simeq \frac{\Omega v}{\sqrt 2}\left ( 1+4(\xi-\xi_2)\tanh\xi_2\right)
\end{equation}
This leads to the relation:
\begin{equation}
\delta=\xi_3-\xi_2=\frac{1}{4\tanh\xi_2}\left(\exp\left(4\frac{\Omega \tilde t}{\sqrt 2}v\tanh\xi_2\right)-1\right)
\end{equation}

Matching the final velocity to the de-Sitter space--time, we get the exit velocity of the particle as:

\begin{equation}
U^{\nu(D_{2})}=e^{\nu(D_{2})}_{a}e^{a(G)}_{\mu}U^{\mu(G)f}
\end{equation}
After some simple calculations, the final velocity of the test particle is obtained as: 
\begin{equation}
U^{\mu(D_2)}=\left(\frac{1}{\sqrt{1-\frac{r_3^2}{a^2}}}(\dot t'_3+\sqrt{2}\dot\phi'_3\sinh^2\xi_3),\ \sqrt{1-\frac{r_3^2}{a^2}}\dot r'_3,\ \frac{1}{r_3}\dot\phi'_3\sinh\xi_3\cosh\xi_3,\ 0\right)
\label{eq10}
\end{equation}

\begin{figure}
  \begin{center}
    \includegraphics[height=10cm]{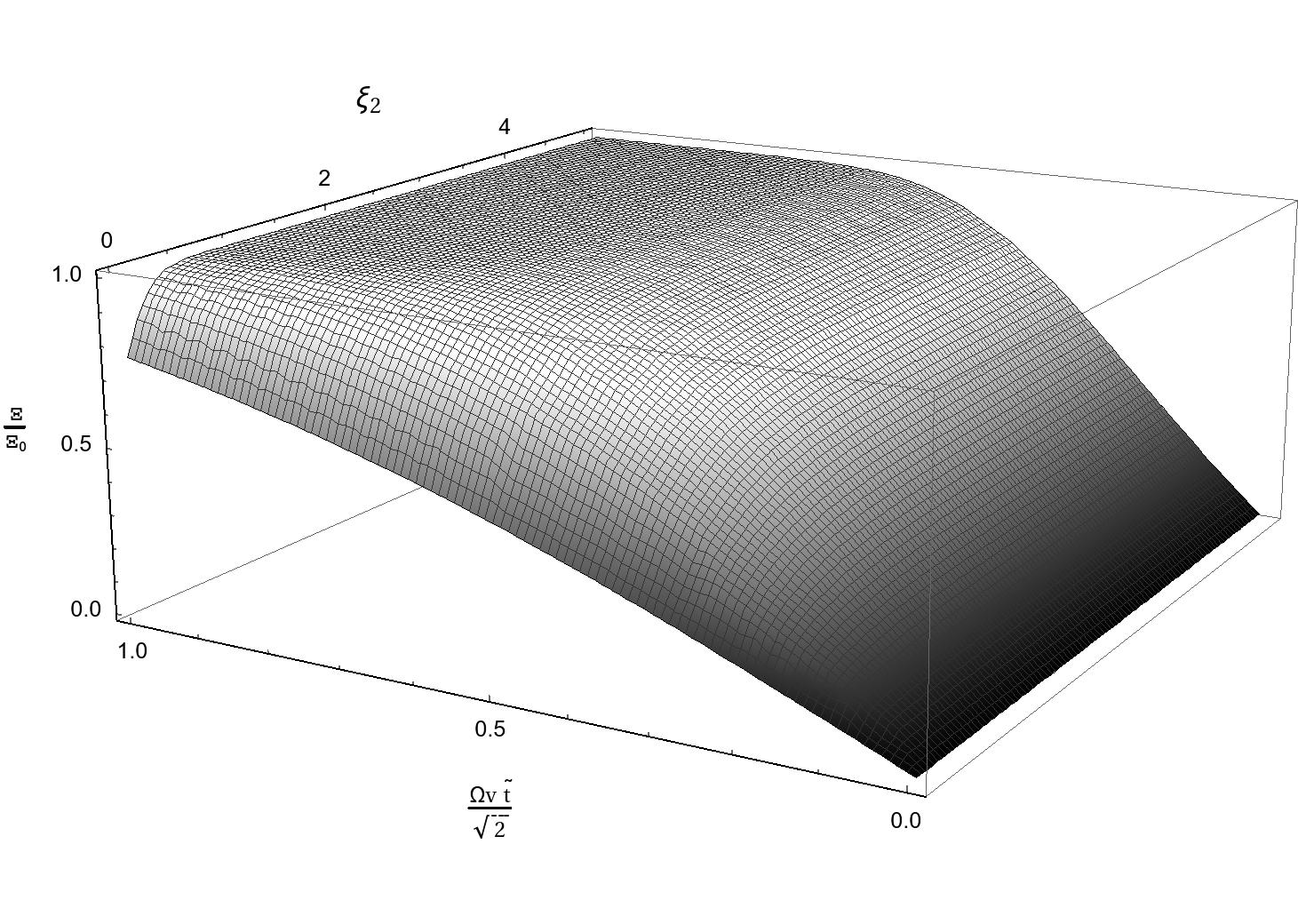}
    \caption{Induced rotation profile as a function of normalized position and initial radial velocity.}
       \label{fig: 5}
  \end{center}
\end{figure}

In order to see how much this particle is rotating, one can evaluate the vorticity: 
\begin{equation}
\Xi =\frac{d\phi }{dt}=\frac{U^2}{U^0}
\end{equation}
Equation (\ref{eq10}) leads to:
\begin{equation}
\frac{\Xi}{\Xi_0} =\frac{\sinh^2(\xi_2+\delta)-\sinh^2\xi_2}{\sinh(\xi_2+\delta)\cosh(\xi_2+\delta)} 
\label{eq11}
\end{equation}
where $\Xi_0=\sqrt{2}\sqrt{1-r_3^2/a^2}/r_3$. The rotation profile is plotted in figure (\ref{fig: 5}). 

Therefore such a sequence of phase transitions induce rotation in the motion of non--rotating particles. The value of this rotation depends on the position of the particle in the de-Sitter space, its distance from the G\"odel's axis of symmetry, and its initial radial velocity but it is of order $\sqrt{\Lambda}$.

After investigating the induced rotation for a test particle, one could ask whether the mechanism works for a congruence of particles or not. 
In what follows we use numerical simulation to investigate the induced rotation  of local and global congruence of particles. 

\begin{figure}
  \begin{center}
    \includegraphics[height=10cm]{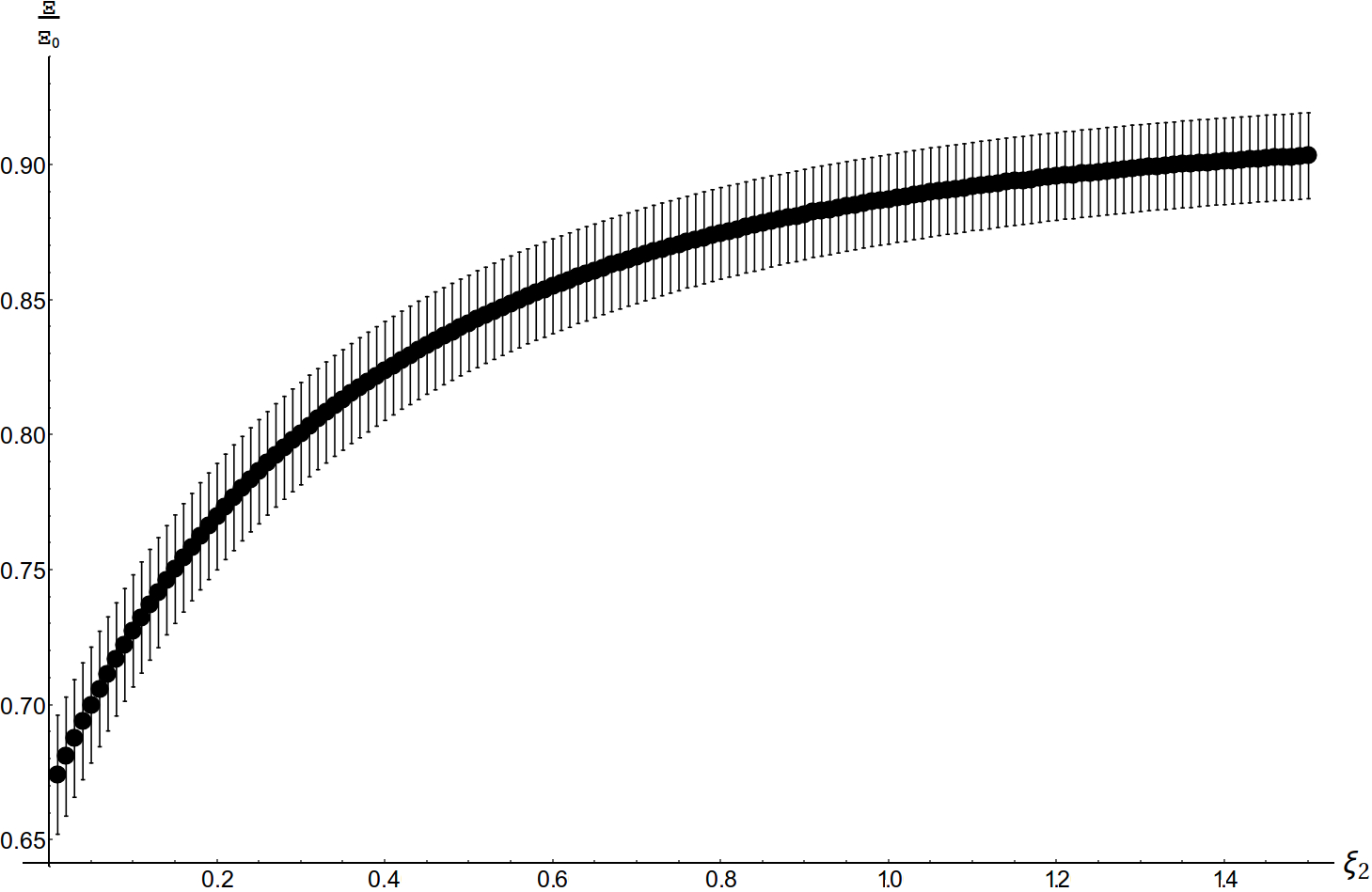}
    \caption{Induced rotation for a local congruence of particles. Each point in this plot shows the result of computer simulation of $10^5$ particles uniformly distributed in the region $(\xi_2,\xi_2+0.01)$. The velocity distribution is assumed to be Gaussian, i.e. $\frac{\Omega v \tilde t}{\sqrt 2}$ is a Gaussian of mean value $0.2$ and width $0.01$. Error bars show the standard deviation of the induced rotation.}
       \label{fig: 6}
  \end{center}
\end{figure}

In order to simulate the local induced rotation, we consider a uniform distribution of a large number (in the order of $10^5$) of 
particles within $\xi_{2}$ and $\xi_{2}+d\xi_{2}$ having a Gaussian distribution in the initial velocity. Figure (\ref{fig: 6}) shows the induced rotation at
different values of $\xi_{2}$ for a local congruence of test particles. This shows that if the phase transition takes place, one have non--zero local  induced rotation.

\begin{figure}
  \begin{center}
    \includegraphics[height=10cm]{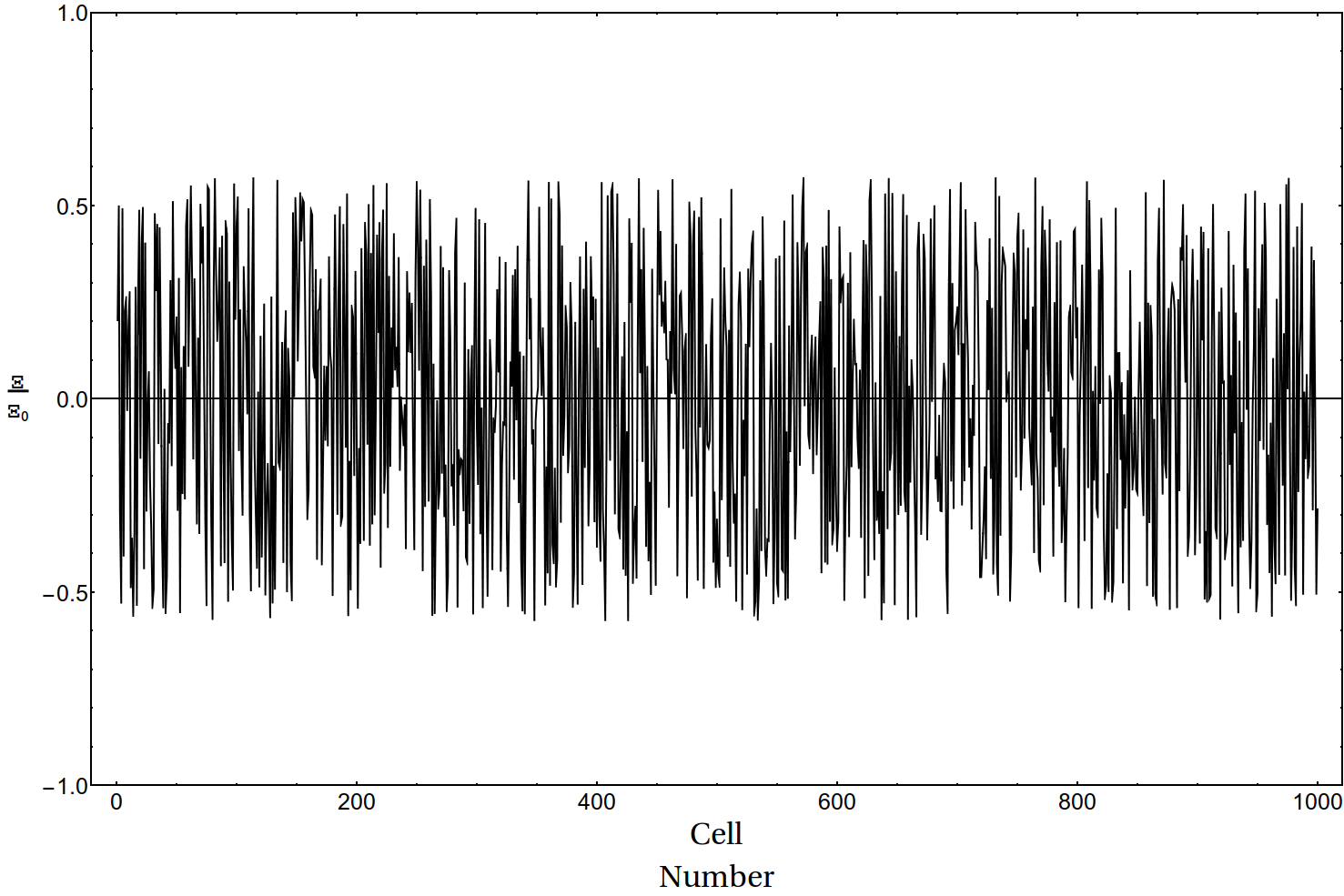}
    \caption{Computer simulation of the global induced rotation. The simulation is done for $1000$ adjacent cells each with $1.5\times 10^5$ particles uniformly distributed in a width $1.5$ in $\xi_2$. The velocity distribution is assumed to be Gaussian, i.e. $\frac{\Omega v \tilde t}{\sqrt 2}$ is a Gaussian of mean value $0.2$ and width $0.01$. The axis of rotation is randomly distributed between cells and the quantum probability of tunneling is considered.}
       \label{fig: 7}
  \end{center}
\end{figure}

On the other hand, to investigate the situation for a global congruence, we divide the space into adjacent cells. Each cell may or may not experience the phase transition according to 
the quantum tunneling probability. The symmetry axis of the G\"odel space is randomly may be any direction. The result of the simulation is shown in figure (\ref{fig: 7}). It is clear that the average global induced rotation is very close to zero. In fact for our simulation the average value of $\Xi/\Xi_0$ is $\sim 0.0045$ with the standard deviation $\sim 0.3312$. This gives a value of $\sim 0.0045\sqrt\Lambda c\sim 1.2\times 10^{-21} \textrm{rad}/\textrm{sec}$ for the global rotation.  Compared to the observational limit\cite{obs}, $\sim 6\times 10^{-21} \textrm{rad}/\textrm{sec}$, this is quite acceptable for our simple model.

As a result, although the local induced rotation is non--zero, the global induced rotation is below the limit of observational data.

\section{Conclusion}
Here we provided a model based on quantum field theory and general relativity which may be a candidate for a possible explanation of the universal rotation. Rotation is essential for the stability of celestial bodies and its origin is an important issue in cosmology. There are number of ways in which physicists have tried to justify how the rotation has been started in the universe, but non of them is the final solution. % to the problem.

We considered a scalar field which may be the inflaton  field and studied a cosmological phase transition between de-Sitter and G\"odel space-times.
We saw that it is quite possible to have a sequence of phase transitions from de-Sitter to G\"odel and then again to de-Sitter space--time. As the universe expands, it reaches a critical temperature in which the potential shows a negative valued minimum. 

If this happens at right time, this minimum which acts as a negative cosmological constant has the needed relation with the dust density leading to G\"odel space--time. As the space--time transits from de-Sitter to G\"odel, the potential changes shape and then the field slow rolls to the new minimum which acts like a positive cosmological constant. The space--time again would be de-Sitter.

 After investigation of the possibility of desired phase transitions, we studied the motion of test particles during these phase transitions. We observed that a non-rotating particle, leaves the G\"odel space--time with  an induced rotation. The induced rotation is a function of the distance from the symmetry axis of G\"odel's space--time and the initial radial velocity of the particle, and is of order of $\sqrt{\Lambda}c$. Simulation shows that for a local congruence of particles we have a non--zero induced rotation, while the global rotation approximately averages to zero.
 
 It should be noted that one should not worry about the exotic behaviors of G\"odel's space--time, because the closed time-like loops are non-geodesics paths in G\"odel's space--time.

This 
model suggests that such a scenario may be possible in the framework of the standard model of cosmology. 
The model may be extended to other space--times and fields (like vector, spinor and tensor fields)  to get closer to an exact enough expression for the universal rotation.
\\

\appendix
\renewcommand{\theequation}{\arabic{equation}}
\setcounter{equation}{0}
\section{One--loop effective potential of a scalar field in the static de-Sitter background}

Starting from the expression (3.21) of \cite{19}
\[
\zeta(z,\beta;\Delta)=\sum_{k=0}^\infty C_k(z)\Delta^k\left \{ \sum_{m=0}^\infty \left [ {\atop{\ }{\ }}\zeta_R(2z+2k-2,mT+3/2)\right.\right.
\]
\[
\left. -2mT\zeta_R(2z+2k-1,mT+3/2)+\left( m^2T^2-\frac{1}{4}\right) \zeta_R(2z+2k,mT+3/2)\right]
\]
\begin{equation}
\left. -\frac{1}{2}\left[ \zeta_R(2z+2k-2,3/2)-\frac{1}{4}\zeta_R(2z+2k,3/2)\right]\right\}
\end{equation}
in which $T=\beta_H/\beta=2\pi a/\beta$, $\Delta=9/4-a^2V''(\phi)$, and  
\begin{equation}
C_k(z)=\frac{\Gamma(z+k)}{k!\Gamma(z)}
\end{equation}
and using
\begin{equation}
kC_k(z)=zC_{k-1}(z+1)
\end{equation}
we see that
\begin{equation}
\frac{\partial \zeta(z,T;\Delta)}{\partial\Delta}=z\zeta(z+1,T;\Delta)
\end{equation}
and
\begin{equation}
\frac{\partial\zeta'(z,T;\Delta)}{\partial\Delta}=\frac{\partial}{\partial z}\left( z\zeta(z+1,T;\Delta)\right )
\end{equation}
The Hurwitz generalized zeta function is analytic everywhere except at $ z=1 $
\begin{equation}
\lim_{z\rightarrow 1}\zeta_R(z,q)=\frac{1}{z-1}-\psi(q)
\end{equation}
For $T$ not so large; i.e. for $\beta>\beta_H$, we can approximate:
\begin{equation}
\sum_m\longrightarrow \int dm=\frac{1}{T}\int d(mT)
\end{equation}
After some tedious calculations, one arrives at the approximate relation:
\[
\zeta'(0,T;\Delta)\simeq
\] 
\begin{equation}
 (0.62\Delta-0.22\Delta^2)+0.075T+\frac{T^3}{120}+\frac{1}{T}\left( -0.07-0.15\Delta 
+2.7\Delta^2+0.025\Delta^3\right )
\end{equation}
\[+ \frac{\Delta}{T}\left( f_0+f_2+\frac{f_1\Delta}{2}\right) -f_3(T) \]
Where $ f_0 $, $ f_1 $, $ f_2 $ and $ f_3 $ are singular quantities:
\[
f_0=\lim_{A\rightarrow\infty}\left( -A-\frac{A^2}{2}\right.
\]
\begin{equation}
\left .-\frac{H_{A+1/2}}{4} -2 A \ln \Gamma(A+3/2)+2\psi_{-2}(A+3/2)+A^2\psi_0(A+3/2)\right )
\end{equation}
\begin{equation}
f_1=\lim_{A\rightarrow\infty}\left(4\psi_0(A+3/2)\right)
\end{equation}
\begin{equation}
f_2=\lim_{A\rightarrow\infty}\left(2 A \ln\Gamma(A+3/2)-2\psi_{-2}(A+3/2)\right )
\end{equation}
\[
f_3=-2T^3\sum_{n=0}^\infty \frac{B_{2n+4}T^{2n}}{8(n+2)!(2n+1)(2n+3)}\Bigg [ (2n+1)(2n+2)\psi_{2n}(3/2)
\]
\begin{equation}
 -\frac{\psi_{2n+2}(3/2)}{4}\Bigg ]
\end{equation}

Now the renormalized one--loop effective potential can be obtained as:
\begin{equation}
V_{\textrm{eff}}(\phi,T)=V(\phi)-\frac{3T}{16\pi^2 a^4}\left( \zeta'(0,T;\Delta)+\ln (a^2V''(\phi))\right ) +\textrm{Counter Terms}
\end{equation}
Counter terms are introduced to renormalize the potential. Infinities could be absorbed by defining the physical parameters of the potential at some value of the field. For $ \lambda \phi^{4} $ theory we have
\begin{equation}
 \frac{\partial^{2}V_{\textrm{eff}}}{\partial \phi^{2}}\mid_{\phi_{0}}=M^2
 \end{equation} 
 \[\frac{\partial^{2}V_{\textrm{eff}}}{\partial \phi^{2}}\mid_{\phi_{0}}=\lambda  \]
Finally,  the one--loop renormalized effective potential can be written as:
\[
\lambda a^{4}V_{\textrm{eff}}=\lambda a^{4}V_0-a^{4}\sigma ^{2}(\frac{\lambda }{2}\phi ^{2})+\frac{a^{4}}{6}(\frac{\lambda }{2}\phi ^{2})^{2}
-\frac{1}{3}\frac{9\lambda }{16\pi ^{2}}\Bigg [\frac{T^{4}}{120}+0.075T^{2}+2.13T+\]
\[9.51+(0.22T-2.7)\left (\frac{51-60T^{2}-8T^{4}}{240}-2.25T^{2}\right )
+\frac{\lambda a^2 }{2}\phi ^{2}f(T)-a^{2}\sigma ^{2}f(T)+
\]
\begin{equation}
\log\left (1-a^{2}\sigma ^{2}+\frac{\lambda }{2}a^{2}\phi ^{2}\right)\Bigg ]
\end{equation}
in which
\begin{equation}
f(T)=0.22T^3-2.7T^2-0.95T+4.2
\end{equation}

\end{document}